\documentclass{article}

\begin{document}


\title{Exact models of pure radiation in $R^2$ gravity\\
for spatially homogeneous\\
Shapovalov spacetimes type II}

\author{
Konstantin~Osetrin\thanks{osetrin@tspu.edu.ru},  \\[1ex]
Altair~Filippov\thanks{altair@tspu.edu.ru},  \\[1ex]
Evgeny~Osetrin\thanks{evgeny.osetrin@gmail.com}, \\[2ex]
Tomsk State Pedagogical University, \\[1ex]
Tomsk, 634061, Russia\\
}



\maketitle

\begin{abstract}
Presented are exactly integrable models with pure radiation in $R^2$ gravity with a cosmological constant, related to wave-like \mbox{Shapovalov} spacetimes type II. Spatially homogeneous models of Shapovalov spaces were considered. The obtained solutions belong to spaces of type III according to the Bianchi classification. For the models under consideration, exact solutions for the equations of motion of test particles are obtained in the Hamilton-Jacobi formalism.
\end{abstract}

\section{Introduction}	

The construction of models for the early stages of the of the Universe expansion is often based on the concept of the radiation dominant era when deviations from the homogeneity and isotropy of space-time are possible and it is necessary to take into account quantum corrections to the field equations. One of the most famous models, in this case is, $R^2$ gravity (see f.e. \cite{1}-\cite{Nojiri2017}). Note that due to the complexity of the field equations, there are few exact solutions for this theory, usually these are solutions inherited from Einstein's theory.

In this work, we obtained exact models in $R^2$ gravity, which have a number of interesting properties.

First, the models considered in this work refer to Shapovalov wave-like spaces  (see \cite{4},\cite{5},\cite{Symmetry}), i.e. allow separating the variables in the eikonal equation and the Hamilton-Jacobi equation for test particles, and one of the separating variables is the wave one, which indicates the wave nature of the solutions.

Secondly, the considered spaces are homogeneous, but non-isotropic models of space-time, i.e., are more complex models than the standard Friedman-Robertson-Walker solution and are adequate as models for the early stages of the expansion of the Universe.

In the third, the considered models use pure radiation as the energy-momentum tensor of matter, which can simulate the high-energy component of various types of radiation (gravitational, electromagnetic, dark energy) and corresponds to the concept of the radiation dominant epoch of the Universe expansion at the stage of its isotropization.

As it is known (\cite{1}-\cite{Nojiri2017}), the field equations of $R^2$ gravity are obtained from the action of the form
\begin{equation}
S=\frac{1}{2\kappa^2}\int d^4x\sqrt{-g}\, (2\Lambda+R+\gamma R^2) +S_M
,
\end{equation}
where $\Lambda$ is the cosmology constant, $R$ is the scalar curvature, $\gamma$ is the constant and
$S_M$ is the action of the matter fields.

The field equations of the theory under consideration can be written in the following form:
\begin{equation}
(1+2\gamma R)\, R_{ij}-\frac{1}{2}\left(2\Lambda+R+\gamma R^2\right) g_{ij}
-2\gamma \left( \nabla_i\nabla_j-g_{ij}\nabla^k\nabla_k\right)R=\kappa^2T_{ij}
\label{R2EqField}
,
\end{equation}
where $\nabla_i$ is the covariant derivative, $T_{ij}$ is the energy-momentum tensor of the matter fields.

The energy-momentum tensor of pure radiation has the form
\begin{equation}
T_{ij}=\varepsilon L_i L_j
\qquad
L^kL_k=0
,
\label{EnergyMomentumTensor}
\end{equation}
where $\varepsilon$ is the radiation energy density, $L^i$ is the wave vector.

The paper deals with Shapovalov spaces of type II. According to the definition of Shapovalov spaces (see \cite{4,5,Symmetry}), these spaces admit the privileged coordinate systems, where the eikonal equation and the Hamilton-Jacobi equation of test particles can be integrated by the method of complete separation of variables, and among the non-ignored separated variables, there are wave variables. Therefore, Shapovalov spaces are called wave-like. The type of Shapovalov space indicates the number of ignored variables on which the metric of the space does not depend or, which is the same, the number of mutually commuting Killing vectors admitted by these spaces. 

In this paper, we obtain exact solutions to the field equations of $R^2$ theory of gravity 
(\ref{R2EqField}) for classes of wave-like spatially homogeneous space-time models that allow separation of
isotropic (wave) variables in the eikonal equation
\begin{equation}
g^{ij}\,\partial_i\Psi\,\partial_j\Psi=0
\label{EqEikonal}
\end{equation}
and in the Hamilton-Jacobi equation for a test particle of mass $m$
\begin{equation}
g^{ij}\,\partial_i S\,\partial_j S=m^2,
\label{EqHJ}
\end{equation}
where $ \Psi $ is the eikonal function, $ S $ is the test particle action function, $ \partial_i $ is the partial derivative.

Spaces allowing the integration of the eikonal equation and the Hamilton-Jacobi equation of the test particles by the method of complete separation of variables have found wide application in the theory of gravity and cosmology, including the study of models with an electromagnetic field (\cite{Obukhov1983}, \cite{Obukhov1986}, \cite{Obukhov1988},\cite{Obukhov20202050186},\cite{Obukhov20201289}), dust matter \cite{OsetrinDust2016}, pure radiation (\cite{OsetrinVaidya1996},\cite{OsetrinVaidya2009},\cite{OsetrinRadiation2017}), scalar fields (\cite{OsetrinScalar212020}, \cite{OsetrinScalar312020}, \cite{OsetrinScalar2018}), spinor fields (\cite{Obukhov1992}, \cite{OsetrinSpinor2013}). 

In works (\cite{OsetrinHomog312002}, \cite{OsetrinHomog2006}, \cite{OsetrinHomog312020}, \cite{OsetrinHomog212020} ) we have carried out a classification of these types of spaces from the point of view of the symmetries they admit, ensuring their spatial homogeneity, i.e. selection of types related to cosmological problems.

For Shapovalov spaces of type II (two non-ignored variables in a privileged coordinate system), there are two spatially homogeneous wave-like spacetime models related to the type III according to Bianchi's classification \cite{OsetrinHomog212020}.
For the case of two non-ignored variables, there are two types of wave-like Shapovalov spaces - types
 {B1} and {B2} (see \cite{OsetrinHomog2006}, \cite{OsetrinHomog212020}), which will be discussed below.
 
The methods considered in this paper may be of interest when constructing cosmological models for f(R) theories of gravity and in models with Gauss – Bonnet terms (see f.e. \cite{Ivashchuk2015}, \cite{Ivashchuk2016} ).
 
\section{Spatially homogeneous wave-like model type~{B1} }

The interval for the space-time model of type {B1} has the form (see \cite{OsetrinHomog212020}):
\begin{equation}
ds^2=
\frac{1}{{x^3}^2}
\left(
2\,dx^0 dx^1+({x^0}-{\alpha})^{1-{\beta}} ({x^0}+{\alpha})^{1+{\beta}}\,{dx^2}^2+{dx^3}^2
\right),
\label{MetricSubB1}
\end{equation}
where $ x^0 $ is an isotropic (wave) variable, $\alpha$ and $\beta$ are constants ($\alpha\ne 0$, $\beta\ne\pm 1$).
$$
{g=\det g_{ij}=}-\frac{({x^0}-{\alpha})^{1-{\beta}} ({\alpha}+{x^0})^{{\beta}+1}}{{x^3}^8},
\qquad
x^0>|\alpha|.
$$
%
%
Independent Killing vector fields of model type {B1} in a privileged coordinate system can be selected in the form:
\[
X_0=\partial_1,
\qquad
X_1=\partial_2,
\qquad
X_2=2\,x^1\partial_1+ x^2\partial_2+x^3\partial_3,
\]
\begin{equation}
X_3= ({x^0}^2-\alpha^2)\partial_0-\frac{{x^3}^2}{2}\partial_1+ {\alpha} \beta x^2\partial_2+x^0 x^3\partial_3.
\end{equation}
Killing vectors $ X_1 $, $ X_2 $, $ X_3 $ define a subgroup of spatial homogeneity of the model.
Killing vector commutators of  model type {B1} have the form:
\[
[{X_0},{X_1}]=0,
\qquad
[{X_0},{X_2}]=2{X_0},
\qquad
[{X_0},{X_3}]=0,
\]
\begin{equation}
[{X_1},{X_2}]={X_1},
\qquad
[{X_1},{X_3}]={\alpha}\beta{}{X_1},
\qquad
 [{X_2},{X_3}]=0.
\end{equation}
For $ \alpha\beta = 0 $, this space admits a third commuting Killing vector and degenerates into
a space with one non-ignored variable only in the privileged coordinate system.

The Riemann tensor $ R_{ijkl} $, the Ricci tensor $ R_{ij} $ and the scalar curvature~$R$ have the following nonzero components:
\begin{equation}  
R_{0313}=-\frac 1{{x^3}^4},
\quad
R_{0212}=R_{2323}=-\frac{(x^0-\alpha)^{1-\beta}(x^0+\alpha)^{1+\beta}}{{x^3}^4}, 
\end{equation} 
\begin{equation}  
R_{0202}=\frac{\alpha^2(1-\beta^2)(x^0-\alpha)^{-1-\beta}(x^0+\alpha)^{-1+\beta}}{{x^3}^2},
\end{equation} 
\begin{equation}  
R_{00}=\frac{\alpha^2(1-\beta^2)}{({x^0}^2-\alpha^2)^2},
\qquad
R_{01}=R_{33}=-\frac 3{{x^3}^2},
\end{equation} 
\begin{equation}  
R_{22}=-\frac{3\,(x^0-\alpha)^{1-\beta}(x^0+\alpha)^{1+\beta}
}{{x^3}^2},
\qquad
R=-12
.
\end{equation} 
Substituting the metric (\ref{MetricSubB1}) into the field equations (\ref{R2EqField}) with the energy-momentum tensor of pure radiation (\ref{EnergyMomentumTensor}), we obtain the following results:
\begin{equation}
\Lambda=3,
\qquad
L_k=\left(L_0,0,0,0\right),
\qquad
L^k=(0,L^1,0,0),
\label{B1cond1}
\end{equation}
and for the radiation we obtain the condition
\begin{equation}
{\varepsilon}\kappa^2 {L^1}^2=
{\alpha}^2 \left(1-{\beta}^2\right)\left(1-24 {\gamma}\right)
\left(
\frac{ {x^3}^2}{{x^0}^2-{\alpha}^2}
\right)^2
.
\label{B1cond2}
\end{equation}
If we assume the radiation energy density  ${\varepsilon}$ and the factor $(1-24 {\gamma})$ to be positive, then we obtain restrictions on the values of the constant $\beta$:
\begin{equation}
-1<\beta<1.
\label{B1cond3}
\end{equation}

Thus, the metric (\ref{MetricSubB1}) and the conditions (\ref{B1cond1}) -- (\ref{B1cond3}) give us an exact solution to the field equations of $ R^2 $ gravity with the energy-momentum tensor of pure radiation.

If $ \alpha \ne 0 $ and $ \beta \ne \pm 1 $ the resulting solution cannot be conformally flat, since the two components of the Weyl tensor $C_{ijkl}$ are not equal to zero:
\begin{equation}
C_{0202} = \frac{{\alpha}^2 \left(1-{\beta}^2\right) ({x^0}-{\alpha})^{-{\beta}-1} ({x^0}+{\alpha})^{{\beta}-1}}{2 {x^3}^2},
\end{equation}
\begin{equation}
C_{0303} = \frac{-{\alpha}^2 \left(1-{\beta}^2\right)}{2 {x^3}^2 ({x^0}-\alpha)^2 ({x^0}+{\alpha})^2}
.
\end{equation}
If  $ {\alpha} = 0 $ or $\beta=0, \pm 1$, the metric of the model type~{B1} degenerates -
in a privileged coordinate system it depends on one variable only.

The model {B1} is of type {III} according to the Bianchi classification and has the type N according to the Petrov classification.

\subsection{Solution of  the Hamilton-Jacobi equation of test particle for the model type {B1}}

As mentioned earlier, the model type {B1} admits in the used privileged coordinate system complete separation of variables in the eikonal equation and the Hamilton-Jacobi equation for test particles. The separation constants $\lambda_1$, $\lambda_2$ and $\lambda_3$ are integrals of motion and are determined by the initial data.

From the Hamilton-Jacobi equation (\ref{EqHJ}) by the method of complete separation of variables we have
the function $ S $ of the action of the test particle for model type {B1} in the form ($\alpha\beta\lambda_1\ne 0$):
$$
S=
\frac{\lambda_3}{2 {\lambda_1}}\,x^0-
\frac{{\lambda_2}^2 }{2\alpha\beta {\lambda_1}} ({x^0}-{\alpha})^{\beta} ({x^0}+{\alpha})^{-\beta}  
+\lambda_1x^1+\lambda_2x^2
+m \log x^3
$$
\begin{equation}
+\sqrt{m^2-\lambda_3 {x^3}^2}
-m \log \left(m+\sqrt{m^2-\lambda_3 {x^3}^2}\right)
+F(\lambda_1,\lambda_2,\lambda_3)
,
\end{equation}
where $\lambda_1$, $\lambda_2$ and $\lambda_3$  are the independent constants of the motion of  test particles and  $F(\lambda_1,\lambda_2,\lambda_3)$ is an arbitrary function of parameters.

Using the proper time of the test particle $\tau=S/m$ (we put the mass of the test particle m = 1)
$$
\tau=
\frac{{\lambda_3} {x^0}}{2
   {\lambda_1}}
-\frac{{\lambda_2}^2 ({x^0}-\alpha )^{\beta } (\alpha +{x^0})^{-\beta }}{2
   \alpha  \beta  {\lambda_1}}+{\lambda_1} {x^1}+{\lambda_2}   {x^2}
$$
\begin{equation}
\mbox{}
+\sqrt{1-{\lambda_3} {x^3}^2}+\log ({x^3})-\log \left(\sqrt{1-{\lambda_3}
   {x^3}^2}+1\right)
\label{tau}
,
\end{equation}
one can obtain a parametric dependence of the coordinates of a particle on its proper time.

According to the Hamilton-Jacobi method, the trajectory of a particle is determined by relations of the form:
\begin{eqnarray}
&&
\frac{\partial S}{\partial \lambda_1} = 
\frac{{\lambda_2}^2 ({x^0}-\alpha )^{\beta } (\alpha +{x^0})^{-\beta }}{2
   \alpha  \beta  {\lambda_1}^2}
  -\frac{{\lambda_3} {x^0}}{2
   {\lambda_1}^2}+{x^1} +{c_1}=0
,
\label{Eq1} \\
&&
\frac{\partial S}{\partial \lambda_2} = 
-\frac{{\lambda_2} ({x^0}-\alpha )^{\beta } (\alpha +{x^0})^{-\beta }}{\alpha 
   \beta  {\lambda_1}}+{x^2}+{c_2}=0
,
\label{Eq2}\\
&&
\frac{\partial S}{\partial \lambda_3} = 
\frac{{x^0}}{2 {\lambda_1}}
-\frac{{x^3}^2}{2   \left(m+\sqrt{m^2-{\lambda_3} {x^3}^2}\right)}
+{c_3}=0
.
\label{Eq3}
\end{eqnarray}
where $c_k=\partial F/\partial\lambda_k$ are additional constants.

If $\lambda_3<0$, then the motion of the test particle is infinite. 
For $\lambda_3>0$ the motion of the test particle along the coordinate $x^0$ and $x^3$ is finite
 and there are turning points.

The turning points of the motion of test particles along the coordinate $x^0$ for $\lambda_3>0$ are determined by solutions of the equation
$$
 {{\lambda_2}}{}^2\left[ ({x^0}-{\alpha})/ ({x^0}+{\alpha})\right]^{\beta} =
 \lambda_3 \left({x^0}^2-\alpha^2\right)
.
$$
For example, for $\lambda_3>0$ and $\beta=0$ we get
$$
|\alpha|<x^0 \le \sqrt{\alpha^2+{\lambda_2}^2/\lambda_3}
.
$$
There are also for $\lambda_3>0$ the turning points  along the coordinate $x^3$:
$$
0< \left|x^3\right| \le m/\sqrt{\lambda_3}
.
$$
Points $x^0=|\alpha|$ and $x^3=0$ are special because the model has a singularity at these points.

Let us choose the variable $x^0$ as a parameter 
$x^0=t$, then from (\ref{Eq1}) we have $x^1=x^1(t)$, from (\ref{Eq2})  we get  $x^2=x^2(t)$, from (\ref{Eq3})  $x^3=x^3(t)$ and from (\ref{tau})  $\tau=\tau \bigl( t,x^1(t),x^2(t),x^3(t)\bigr)=\tau(t)$.
Thus, we have determined the parametric dependence of the coordinates of the particle as functions of its proper time $\tau$.

The synchronous coordinate system associated with the test particle, where the test particle is at rest, is determined by transformation from $x^k$ to $\tilde x^k=\left(\tau,\lambda_1,\lambda_2,\lambda_3\right)$, given by equations (\ref{tau}), (\ref{Eq1}), (\ref{Eq2}), (\ref{Eq3}).

The eikonal function $\Psi$ from Eq.(\ref{EqEikonal}) for the model type {B1} has the form:
\begin{equation}
\Psi=
\frac{{\lambda_3}^2}{2 {\lambda_1}}\,x^0-
\frac{{\lambda_2}^2 }{2\alpha\beta {\lambda_1}}
\left( \frac{{x^0}-{\alpha}}{{x^0}+{\alpha}} \right)^{\beta}  
+\lambda_1x^1+\lambda_2x^2
+\lambda_3 x^3
+F(\lambda_1,\lambda_2,\lambda_3)
,
\end{equation}
where $\lambda_1$, $\lambda_2$, and $\lambda_3$  are the independent constants and  $F(\lambda_1,\lambda_2,\lambda_3)$ is an arbitrary function of parameters.

\section{Spatially homogeneous wave-like model  type~{B2}}

The space-time interval for a  type~{B2} model  can be written as (see \cite{OsetrinHomog212020}):
\begin{equation}
ds^2=\frac{1}{{x^3}^2}\,\left(
2\,dx^0dx^1+{x^0}^{\alpha}\,{dx^2}^2 +{dx^3}^2
\right),
\label{MetricSubB2}
\end{equation}
where $ x^0 $ is an isotropic (wave) variable, and ${\alpha}$ is a constant.
$$
{g=\det g_{ij}=}-{x^0}^{\alpha}/{x^3}^8,
\qquad
x^0>0
.
$$
Independent Killing vector fields in a privileged coordinate system can be selected in the form:
\[
X_0=\partial_1,
\qquad
X_1=\partial_2,
\qquad
X_2=2\,x^1 \partial_1+x^2\partial_2+x^3\partial_3,
\]
\begin{equation}
X_3=x^0\partial_0+ \frac{1-\alpha}{2}\,x^2\partial_2+\frac{x^3}{2}\,\partial_3.
\end{equation}
Killing vectors $ X_1 $, $ X_2 $, $ X_3 $ define a subgroup of spatial homogeneity of the model.
Killing vector commutators of  model  type~{B2} have the form:
\[
[{X_0},{X_1}]=0,
\qquad
[{X_0},{X_2}]=2{X_0},
\qquad
[{X_0},{X_3}]=0,
\]
\begin{equation} 
[{X_1},{X_2}]={X_1},
\qquad
[{X_1},{X_3}]=\frac{1-\alpha}{2}\,{X_1},
\qquad
[{X_2},{X_3}]=0.
\end{equation}
For $ {\alpha} =  1 $, this space admits an additional  commuting Killing vector and degenerates into a space with one non-ignored variable.

The Riemann tensor $ R_{ijkl} $, the Ricci tensor $ R_{ij} $ and the scalar curvature $ R $ have the following nonzero components:
\[
R_{0101}=-R_{0313}=\frac 1{{x^3}^4},\ \ \ R_{0212}=R_{2323}=-\frac{{x^0}^{\alpha}}{{x^3}^4},\ \ \ 
R_{0202}=\frac{\alpha(2-\alpha)}{4{x^3}^2{x^0}^{(2-\alpha)}}, 
\]
\begin{equation} 
 R_{01}=R_{33}=-\frac 3{{x^3}^2},\ \ \ R_{22}=-\frac{3{x^0}^{\alpha}}{{x^3}^2},\ \ \ R_{00}=\frac{\alpha(2-\alpha)}{4{x^0}^2},\ \ \ R=-12
 .
 \end{equation} 
Substituting the metric (\ref{MetricSubB2}) into the field equations (\ref{R2EqField}) with the energy-momentum tensor of pure radiation (\ref{EnergyMomentumTensor}), we obtain the following results:
\begin{equation} 
\Lambda=3,
\qquad
L_k=\left(L_0,0,0,0\right),
\quad
L^k=\left( 0, L^1,0,0,0\right),
\quad
L^1={x^3}^2\, L_0,
\label{B2cond1}
\end{equation} 
\begin{equation} 
\kappa  {\varepsilon} {L^1}^2=\frac{{\alpha} (2-{\alpha})(1-24 {\gamma})\, {x^3}^4 }{4\,{x^0}^2}
.
\label{B2cond2}
\end{equation} 
If we assume the radiation energy density  ${\varepsilon}$ and the factor $(1-24 {\gamma})$ to be positive, then we obtain restrictions on the values of the constant $\alpha$:
\begin{equation}
0\le\alpha\le 2,
\label{B2cond3}
\end{equation}
Thus, the metric (\ref{MetricSubB2}) and the conditions (\ref{B2cond1}) -- (\ref{B2cond3}) give us an exact solution of the field equations of $ R^2 $ gravity with the energy-momentum tensor of pure radiation.

If $ \alpha \ne 0 $ or $ \alpha \ne 2 $  the resulting solution cannot be conformally flat, since the two components of the Weyl tensor $C_{ijkl}$ are not equal to zero:
\begin{equation} 
{\rm C}_{0202} =\frac{{\alpha}\, (2-{\alpha}) {x^0}^{({\alpha}-2)}}{8 {x^3}^2},
\qquad
{\rm C}_{0303} = -\frac{{\alpha}\, (2-{\alpha})}{8 {x^0}^2 {x^3}^2}
 \end{equation} 
If $ \alpha = 0 $ or $ \alpha = 2 $, the Weyl tensor vanishes (conformally flat space), but the Ricci tensor, scalar curvature and the Riemann curvature tensor does not vanish. 
If  $ {\alpha} = 0, 1, 2 $, the metric of the model type~{B2} degenerates -
in a privileged coordinate system it depends on one variable only.

This spatially homogeneous space-time model is of type {III} according to the Bianchi classification and has type N according to the Petrov classification.

\subsection{Solution of  the Hamilton-Jacobi equation of test particle for the model type~{B2}}

As mentioned earlier, the model type {B2} admits in the privileged coordinate system complete separation of variables in the eikonal equation and the Hamilton-Jacobi equation for test particles. The separation constants $\lambda_1$, $\lambda_2$, and $\lambda_3$ are integrals of motion and are determined by the initial data.

From the Hamilton-Jacobi equation (\ref{EqHJ}) by the method of complete separation of variables we have
the function $ S $ of the action of the test particle for the~model type {B2} in the form ($\lambda_1\ne 0$, $\alpha\ne 1$):
$$
S=
\frac{x^0}{2{\lambda_1}}
\left(
{\lambda_3}  +{\lambda_2}^2{x^0}^{-\alpha}/(\alpha-1) 
\right)
+{\lambda_1}x^1+{\lambda_2}x^2 
+m\log x^3
$$
\begin{equation}
\mbox{}+
\sqrt{m^2-{\lambda_3}{x^3}^2}-m\log{\left(m+\sqrt{m^2-{\lambda_3}{x^3}^2}\right)}
+F(\lambda_1,\lambda_2,\lambda_3)
,
\label{B2S}
\end{equation}
where $\lambda_1$, $\lambda_2$, and $\lambda_3$  are the independent constants of the motion of  test particles and  $F(\lambda_1,\lambda_2,\lambda_3)$ is an arbitrary function of parameters.


According to the Hamilton-Jacobi method, the trajectory of a test particle is determined by relations of the form:
\begin{eqnarray}
& &
\frac{\partial S}{\partial \lambda_1}=
x^1-\frac{x^0}{2{\lambda_1}^2}\left( 
{\lambda_3}+{\lambda_2}^2{ x^0}^{-\alpha}/(\alpha-1)
\right)+c_1=0,
\label{B2eq1}
\\
& &
\frac{\partial S}{\partial \lambda_2}=
x^2
+\frac{\lambda_2}{{\lambda_1}(\alpha - 1)} { x^0}^{(1-\alpha)}+c_2=0,
\label{B2eq2}
\\
& &
\frac{\partial S}{\partial {\lambda_3}}=
\frac{x^0}{2{\lambda_1}}
-\frac{{x^3}^2}{2\left(  m+\sqrt{m^2-\lambda_3 {x^3}^2 }\right)}
+c_3=0,
\label{B2eq3}
\end{eqnarray}
where $c_k=\partial F/\partial\lambda_k$ are constants.

The proper time of a test particle $\tau=S/m$ can be written as ($m\to 1$):
$$
\tau=\frac{1}{2{\lambda_1}}\left({\lambda_3} x^0-{\lambda_2}^2\log x^0 \right)+{\lambda_1}x^1+{\lambda_2}x^2 
$$
\begin{equation} 
\mbox{}
+
\sqrt{1-{\lambda_3}{x^3}^2}+\log x^3-\log{\left(1+\sqrt{1-{\lambda_3}{x^3}^2}\right)},
\label{B2tau}
 \end{equation} 
Then, using the variable $x^0=t$ as a parameter, from equations (\ref{B2eq1})-(\ref{B2tau}) we obtain the parametric dependence of the coordinates of the test particle $x^k(t)$  on its proper time $\tau(t)=\tau\left( t,x^1(t),x^2(t),x^3(t)\right)$.

If $\lambda_3<0$, then the motion of the test particle  for model type {B2}  is infinite. 
For $\lambda_3>0$ the motion of the test particle along the coordinate $x^0$ and $x^3$ is finite
 and there are turning points:
\begin{equation} 
0<x^0 \le \left( {\lambda_2}^2/\lambda_3\right)^{1/\alpha},
\qquad
0< \left| x^3 \right| \le m/\sqrt{\lambda_3}.
\end{equation} 
The eikonal function $\Psi$ from Eq.(\ref{EqEikonal}) for the model type {B2} has the following form ($\lambda_1\ne 0$, $\alpha\ne 1$):
\begin{equation}
\Psi=
\frac{x^0}{2{\lambda_1}}
\left(
{\lambda_3}^2  +{\lambda_2}^2{x^0}^{-\alpha}/(\alpha-1) 
\right)
+{\lambda_1}x^1+{\lambda_2}x^2 
+{\lambda_3}{x^3}
+F(\lambda_1,\lambda_2,\lambda_3)
,
\label{B2Psi}
\end{equation}
where $\lambda_1$, $\lambda_2$, and $\lambda_3$  are the independent constants and  $F(\lambda_1,\lambda_2,\lambda_3)$ is an arbitrary function of parameters.

\section{Conclusion}

Two classes of exact solutions are obtained in $R^2$ gravity with a cosmological constant and pure radiation for spatially homogeneous cosmological models allowing the integration of the equations of motion of test particles by the method of separation of variables in the Hamilton-Jacobi formalism. 

The solutions belong to the class II of Shapovalov wave-like spaces and have a separating wave variable in a privileged coordinate system, which reflects the wave nature of the solutions obtained. The positiveness of the energy density of pure radiation imposes restrictions on the parameters of the considered spacetime models. 

For the obtained models, the equations of motion of test particles were integrated in the Hamilton-Jacobi formalism. Complete integrals are presented for the action function of test particles and the eikonal function. Parametric dependence of the coordinates of test particles as functions of proper time is obtained. The solutions obtained make it possible to go over to a synchronous coordinate system with respect to which the freely falling observer is at rest. 

The  obtained models belong to type III according to Bianchi's classification and to type N according to Petrov's classification.

Wave-like spatially homogeneous models of spacetime can describe aperiodic primordial gravitational waves of the Universe.

\section*{Acknowledgments}

The reported study was funded by RFBR, project number N~20-01-00389~A.


\begin{thebibliography}{99}

\bibitem{1} 
S.~Capozziello, M.~De~Laurentis, 
Extended Theories of Gravity, Physics Reports, 509 (4-5), pp.~167-321,
DOI: 10.1016/j.physrep.2011.09.003

\bibitem{2} S. Nojiri and S.D. Odintsov, Introduction to modified gravity and gravitational alternative for dark energy, Int.J.Geom.Meth.Mod.Phys. 4 (2007),
115-146.

\bibitem{3} S. Nojiri and S.D. Odintsov, Unified cosmic history in modified gravity:
from F(R) theory to Lorentz non-invariant models, Physics Reports, 505 (2011),
59-144.

\bibitem{Nojiri2017}
S.~Nojiri, S.D.~Odintsov and   V.K.~Oikonomou,
Modified gravity theories on a nutshell: Inflation, bounce and late-time evolution,
Physics Reports (2017), 692, pp.~1-104,
DOI: 10.1016/j.physrep.2017.06.001

\bibitem{4} V.N. Shapovalov, Symmetry and separation of variables in Hamilton-Jacobi equation, Izv. vuzov. Fizika (Sov. Phys. Journ.) 9 (1978) 18.

\bibitem{5}  V.N. Shapovalov, The St\"{a}ckel spaces, Sib. Math. Journal (Sov. J. of Math.) 20 (1979) 1117.

\bibitem{Symmetry}
K.~Osetrin, E.~Osetrin,
Shapovalov wave-like spacetimes
(2020) Symmetry, 12 (8),  1372. 
DOI: 10.3390/SYM12081372

\bibitem{Obukhov1983}
V.G.~Bagrov, V.V~Obukhov,
Classes of Exact Solutions of the Einstein‐Maxwell Equations
(1983) Annalen der Physik, 495 (4-5), pp.~181-188. 
DOI: 10.1002/andp.19834950402

\bibitem{Obukhov1986}
V.G.~Bagrov, V.V~Obukhov, A.V.~Shapovalov,
Special St\"{a}ckel electrovac spacetimes
(1986) Pramana, 26 (2), pp.~93-108. 
DOI: 10.1007/BF02847629

\bibitem{Obukhov1988}
V.G.~Bagrov, V.V.~Obukhov, K.E.~Osetrin,
Classification of null-St\"{a}ckel electrovac metrics with cosmological constant
(1988) General Relativity and Gravitation, 20 (11), pp.~1141-1154. 
DOI: 10.1007/BF00758935

\bibitem{Obukhov20202050186}
V.V.~Obukhov, 
Separation of variables in Hamilton--Jacobi equation for a charged test particle in the St\"{a}ckel  spaces of type (2.1), International Journal of Geometric Methods in Modern Physics (2020), 17(14), 2050186.
DOI: 10.1142/S0219887820501868

\bibitem{Obukhov20201289}
V.V.~Obukhov, 
Hamilton--Jacobi equation for a charged test particle in the St\"{a}ckel  spaces of type (2.0),
Symmetry (2020), 12 (8), 1289.
DOI: 10.3390/sym12081289

\bibitem{OsetrinDust2016}  
K.~Osetrin, A~Filippov and E.~Osetrin,
The spacetime models with dust matter that admit separation of variables in Hamilton-Jacobi equations of a test particle. {\it Mod. Phys. Lett.}  {\bf A31} (2016), N~06, 1650027.

\bibitem{OsetrinVaidya1996}
V.G.~Bagrov, A.D.~Istomin, V.V.~Obukhov and K.E.~Osetrin,
Classification of conformal St\"{a}ckel spaces in the Vaidya problem
(1996) Russian Physics Journal, 39 (8), pp.~744-749. 
DOI: 10.1007/BF02437084

\bibitem{OsetrinVaidya2009}
V.V~Obukhov, K.E.~Osetrin, A.E.~Filippov, Y.A.~Rybalov,
The Vaidya problem in conformally flat Stäckel spaces of type (1.1)
(2009) Russian Physics Journal, 52 (1), pp.~11-14. 
DOI: 10.1007/s11182-009-9198-3

\bibitem{OsetrinRadiation2017} 
E.~Osetrin and K.~Osetrin, 
Pure radiation in space-time models that admit integration of the eikonal equation by the separation of variables method. {\it J. Math. Phys.}  {\bf 58} (2017), N~11, 112504.

\bibitem{OsetrinScalar212020}
K.~Osetrin, A.~Filippov and E.~Osetrin,
Wave-like spatially homogeneous models of St\"{a}ckel spacetimes (2.1) type in the scalar-tensor theory of gravity (2020)
Modern Physics Letters A, 2050275.
DOI: 10.1142/S0217732320502752

\bibitem{OsetrinScalar312020}
E.K.~Osetrin, K.E.~Osetrin, A.E.~Filippov and I.V.~Kirnos,
Wave-like spatially homogeneous models of St\"{a}ckel spacetimes (3.1) type in the scalar-tensor theory of gravity (2020)
International Journal of Geometric Methods in Modern Physics,
v 17, N 12, 2050184
DOI: 10.1142/S0219887820501844

\bibitem{OsetrinScalar2018}
K.E.~Osetrin, A.E.~Filippov, E.K.~Osetrin,
Models of Generalized Scalar-Tensor Gravitation Theories with Radiation Allowing the Separation of Variables in the Eikonal Equation (2018) Russian Physics Journal, 61 (8), pp.~1383-1391. 
DOI: 10.1007/s11182-018-1546-8

\bibitem{Obukhov1992}
V.G.~Bagrov, V.V~Obukhov,
New method of integration for the Dirac equation on a curved space-time
(1992) Journal of Mathematical Physics, 33 (6), pp.~2279-2289. 
DOI: 10.1063/1.529600

\bibitem{OsetrinSpinor2013}
K.E.~Osetrin, Y.A.~Rybalov,
Cosmological Models with Scalar and Spinor Fields
(2013) Russian Physics Journal, 55 (12), pp.~1416-1424. 
DOI: 10.1007/s11182-013-9975-x

\bibitem{OsetrinHomog312002}
V.V~Obukhov, K.E.~Osetrin, A.E.~Filippov,
Metrics of homogeneous spaces admitting (3.1)-type complete sets
(2002) Russian Physics Journal, 45 (1), pp.~42-48. 
DOI: 10.1023/A:1016093620137

\bibitem{OsetrinHomog2006}  
V.V.~Obukhov, K.E.~Osetrin, A.E. Filippov, Homogeneous spacetimes and
separation of variables in the Hamilton-Jacobi equation. Journal of Physics~A:
Mathematical and General. 39 (2006), N 21, 6641-6647.

\bibitem{OsetrinHomog312020}
E.K.~Osetrin, K.E.~Osetrin, A.E.~Filippov,
Spatially Homogeneous Conformally St\"{a}ckel Spaces of Type (3.1)
(2020) Russian Physics Journal, 63 (3), pp.~403-409. 
DOI: 10.1007/s11182-020-02050-2

\bibitem{OsetrinHomog212020}  
E.K.~Osetrin, K.E.~Osetrin, A.E.~Filippov,
Spatially Homogeneous Models St\"{a}ckel Spaces of Type (2.1)
(2020) Russian Physics Journal, 63 (3), pp.~410-419. 
DOI: 10.1007/s11182-020-02051-1

\bibitem{Ivashchuk2015}  
V.D.~Ivashchuk,   A.A.~Kobtsev,
On exponential cosmological type solutions in the model with Gauss-Bonnet term and variation of gravitational constant
(2015) European Physical Journal C, 75 (5), 177, pp. 1-12. 
DOI: 10.1140/epjc/s10052-015-3394-9

\bibitem{Ivashchuk2016}  
 V.D.~Ivashchuk,
On stable exponential solutions in Einstein-Gauss-Bonnet cosmology with zero variation of G
(2016) Gravitation and Cosmology, 22 (4), pp. 329-332. 
DOI: 10.1134/S0202289316040095

\end{thebibliography}
\end{document}